\documentclass[final,5p,times,twocolumn,authoryear]{elsarticle}

\usepackage{amssymb}

\usepackage{amsmath,bm,ulem}

\journal{Ocean Modelling}

\begin{document}

\begin{frontmatter}



\title{Optimisation of an idealised ocean model, stochastic parameterisation of sub-grid eddies.}


\author[aopp]{Fenwick C. Cooper\corref{cora}}
\ead{Fenwick@LittleStick.com}

\author[aopp]{Laure Zanna}

\address[aopp]{Atmospheric, Oceanic and Planetary Physics, Department of Physics, University of Oxford, Oxford, OX1 3PU, UK}

\cortext[cora]{Corresponding author. Tel.: +44 7531 290 687}

\begin{abstract}

An optimisation scheme is developed to accurately represent the sub-grid scale forcing of a high dimensional chaotic ocean system. Using a simple parameterisation scheme, the velocity components of a 30km resolution shallow water ocean model are optimised to have the same climatological mean and variance as that of a less viscous 7.5km resolution model. The 5 day lag-covariance is also optimised, leading to a more accurate estimate of the high resolution response to forcing using the low resolution model.

The system considered is an idealised barotropic double gyre that is chaotic at both resolutions. Using the optimisation scheme, we find and apply the constant in time, but spatially varying, forcing term that is equal to the time integrated forcing of the sub-mesoscale eddies. A linear stochastic term, independent of the large-scale flow, with no spatial correlation but a spatially varying amplitude and time scale is used to represent the transient eddies. The climatological mean, variance and 5 day lag-covariance of the velocity from a single high resolution integration is used to provide an optimisation target. No other high resolution statistics are required. Additional programming effort, for example to build a tangent linear or adjoint model, is not required either.

The focus of this paper is on the optimisation scheme and the accuracy of the optimised flow. The method can be applied in future investigations into the physical processes that govern barotropic turbulence and it can perhaps be applied to help understand and correct biases in the mean and variance of a more realistic coarse or eddy-permitting ocean model. The method is complementary to current parameterisations and can be applied at the same time without modification. 

\end{abstract}


\end{frontmatter}



\section{Introduction}

Due to the limitations of finite computational power, current
numerical methods are not capable of accurately resolving the ocean
circulation down to the viscous scale. Since there exists no universal
sub-grid scale turbulence model that can close for all unresolved
quantities (Reynolds stresses, turbulent fluxes, etc.) ad-hoc
representations are required, and state of the art numerical models
exhibit serious differences and inaccuracies in their climatologies
(e.g. \citealp{Flato13}, section 9.4.2). The simplest approach to
parameterise sub-grid scale processes is to dissipate any small-scale
motion while simultaneously stabilising the model. This is typically
achieved by employing an eddy diffusivity designed, for example, to
improve spectral characteristics near the grid-scale
(e.g. \citealp{Smagorinsky-1963:general,Leith67}), or by using a
diffusive integration scheme (e.g. \citealp{Ritchie88}). Another
approach is to mimic the physical processes in the real ocean. For
example, mesoscale eddies in the ocean interior tend to rearrange
fluid parcels along isopycnals (constant density surfaces) which leads
to the widely implemented Gent-McWilliams parameterisation scheme in
the tracer equations \citep{Gent-Mcwilliams-1990:isopycnal}.  Such
approaches to find the sub-grid momentum or buoyancy forcing are often
based upon the time-mean effect of the sub-grid scale forcing upon the
large scale flow as diagnosed by comparing a low resolution model with
measurements, or a high resolution integration. The approximate
functional form of the sub-grid momentum or buoyancy forcing in terms
of the grid scale flow of a turbulent system may be found using high
resolution integrations (e.g. \citealp{Achatz99}), using, for example,
a polynomial fit. A stochastic term may be used to represent the fit
residual
(e.g. \citealp{Wilks05,frederiksenetal2006,Zidikheri09,kitsiosetal2013,Arnold13})
or realistic variance
(e.g. \citealp{Hasselmann-1976:stochastic,Buizza-Miller-Palmer-1999:stochastic,Palmer-2001:nonlinear,berloff2005}). The
deterministic and stochastic sub-grid forcing can be derived from
theoretical considerations
(e.g. \citealp{kraichnan1959,herringetal1972,frederiksenetal1997,Holm99,marshalletal2012,Grooms13,Mana14}),
although such an approach can be practically difficult to implement
(\citealp{Foias01,Mana14}).

In many cases one or more parameters that govern the strength of these
schemes must be chosen with limited guidance from theory. Trial and
error comparison of model output, as a function of parameter values,
with ocean data, is one method often referred to as ``tuning''. Tuning
up to around five or six parameters is possible in principle with a
simple parameter search, however larger numbers of parameters require
an ``expert'' opinion, independence from each other, or an acceptance
that the optimal values will not be found. The problem is that to
explore each direction of a parameter space of dimensionality $d$
across $n$ different climatologies requires $n^d$ points to be
evaluated. Given a high ($d\gg 5$) dimensional vector $\mathbf{p}$ of
parameters evaluation of the entire parameter space is not practical
and in order to optimise anything we are forced to define an objective
target to optimise for, or in other words a cost function
$G(\mathbf{p})$ to minimise. From an initial guess $\mathbf{p}_0$ a
direction to change $\mathbf{p}$ may be given by the gradient of the
cost function
\begin{equation}
  \mathbf{p}_1=\mathbf{p}_0 - \left. \frac{\partial G(\mathbf{p})}{\partial \mathbf{p}} \right|_{\mathbf{p}=\mathbf{p}_0} \delta p.
\label{costFuncMin:eqn}
\end{equation}
Here $\mathbf{p}_1$ is an improved estimate of the optimal parameters
in comparison with $\mathbf{p}_0$ and $\delta p$ is a small positive
constant with the appropriate units. The process can be iterated until
no further optimisation is possible. Accurate estimation of $\partial
G/\partial \mathbf{p}$ can be difficult, requiring for example the use
of tangent linear and adjoint models to be integrated. Implementation
of adjoint models for ocean circulation problems has been achieved for
sensitivity analysis and data assimilation capabilities
\citep[e.g.,][]{Marotzke-Giering-Zhang-et-al-1999:construction,Moore-Arango-Di-et-al-2004:comprehensive}
as has optimisation of the eddy-buoyancy sub-grid parameters from the
climatological mean state \citep{ferreiraetal2005}.  However there are
still some unresolved issues for large-scale chaotic systems. Firstly
the programming effort is substantial leading to the development of
semi-automatic differentiation packages for this purpose
(e.g. \citealp{Giering-1999:tangent,Heimbach-Hill-Giering-2005:efficient}). Secondly
if a system has a stochastic element the problem of optimising
stochastic parameterisation has not, to the author's knowledge, been
considered. Finally, although the adjoint approach is useful for short
time optimisation in ocean
(e.g. \citealp{Gebbie06,Mazloff10,Balmaseda13}) and atmosphere
(e.g. \citealp{Kalnay96,Dee11}) state estimation, it is not currently
capable of optimising for the long
time climate averages of a chaotic system
(e.g. \citealp{Lea-Allen-Haine-2000:sensitivity,Eyink04}) and
approximations are required. Some attempts to solve this problem in a
slightly different context include the methods of \cite{Abramov09}
applied to climate response, who use the full non-linear model for the
short time gradient estimate and a Gaussian model approximation for
longer times, and \cite{Wang14} who uses a modified adjoint algorithm
to stabilise the gradient estimation algorithm. Fortunately an
estimate of $\partial G/\partial \mathbf{p}$ does not need to be
particularly accurate for the purposes of optimisation. It is merely
required to follow a trajectory in parameter space that eventually
leads toward the optimum and to tend to zero as the optimum is
approached. Therefore we have the opportunity to optimise with a much
simpler criteria if a very approximate direction of $\partial
G/\partial \mathbf{p}$ can be found. This is the approach of the
present paper. In our case, with the climate change problem in mind,
the goal is accurate optimisation of the climatological mean and
variance and approximate optimisation of the response of the system to
a forcing, using a ``truth'' as the optimisation target.

\subsection{The mean}
\label{mean:sect}

Current state of the art ocean models exhibit a different
climatological mean state to that observed in the real ocean
\citep{Flato13}. For example, the poor representation of eddy-mean
flow processes leads to unrealistic western boundary currents (Gulf
Stream and Kuroshio) responsible for large sea surface temperature
biases \citep{large06}.  Their predictions are therefore
approximations about a different climatological mean point in state
space to that of reality. To account for such deviations from the
observed climatology, post integration bias correction is sometimes
applied (e.g. \citealp{Stockdale97}).  A more accurate approach would
be to have a model that has the correct climatological mean state in
the first place.  This can be achieved for example by adding a
spatially varying, but constant in time, parameter to the right hand
side of the governing equations \citep{Achatz99}. This
spatially-varying time-independent parameter represents the
contribution to the climatological mean of all of the sub-grid
processes that are not included in the basic low resolution model
minus any biases introduced by
incorrect additional terms, such as high viscosity. The size of the
improvement in accuracy relative to post integration bias correction
can be important. For example, in a coupled ocean-atmosphere model
some studies suggest that the mean location of the ocean boundary
currents have an important impact upon atmospheric dynamics
\citep[e.g.,][]{Kirtman11,Scaife11}. The ocean bias therefore has the
potential to cause atmospheric bias that may be difficult to correct
post integration.

\subsection{The variance}
\label{variance:sect}

Often, due to artificially high viscosity in a dynamical ocean model
and the lack of sub-grid variability, the variance of the prognostic
variables is underestimated.  Without a time dependent external
forcing such as the seasonal cycle, one can often obtain a steady
state in very low-resolution ocean models, where time derivatives of
all prognostic variables are equal to zero. In non-eddying ocean
models, any effect of the variance due to eddies is therefore reduced
or missing.  The fluctuations brought about by resolving the eddies in
an ocean model can potentially lead to additional dynamical regimes
being explored (e.g. \citealp{Palmer-2001:nonlinear,Palmer11}) and
important processes such as eddy saturation
\citep{straub1993,mundayetal2013} or jet rectification
\citep{berloff2005,Waterman-Jayne-2012,watermanetal2013}. In addition,
the lack of variance between the members of an ensemble of model
integrations contributes to over confidence, in a statistical sense,
in model predictions. For these reasons we consider it desirable for
our model climatological variance, and hence the turbulent eddy
kinetic energy, to be as close as possible to the measured ocean
variance. Moreover, since the correlations of a turbulent system decay
in time, we would like the correlations of any parameterised source of
variance to also decay after some time.  The simplest approach is to
add a stochastic term, with a spatially varying amplitude and time
scale, to the right hand side of the governing equations. In this
paper we require that the parameters governing such a process ensure
that the model's climatological variance is as accurate as possible,
relative to the ``truth''.

\subsection{The response to forcing}
\label{ForcedRespIntro:sect}

For climate change experiments or for seasonal forecasts, the accuracy
of a model's response to forcing is an important
characteristic. Unfortunately, unlike in the case of the mean and the
variance, we can not always directly measure the true response of a
system to a particular forcing and compare it with our low resolution
model response. However we would like to be able to predict in advance
the change in the climatological mean and variance due to forcing. A
forcing change may be, for example, an increase in the concentration
of carbon dioxide or in wind stress over the Southern Ocean
(e.g. \citealp{fyfe07}).

\begin{figure}
\begin{center}
\includegraphics[width=0.49\textwidth]{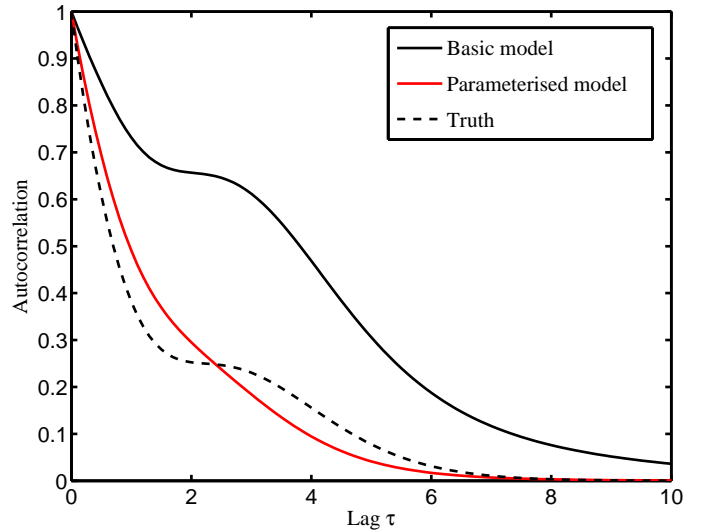}
\end{center}
\caption{In this hypothetical illustration, the integral of the
  autocorrelation function of a model without parameterisation, is
  much larger than the integral of the autocorrelation function of the
  true system. Therefore we might expect that the model's
  climatological response to forcing is somewhat greater than the true
  climatological response to forcing. After adding a linear stochastic
  parameterisation with an appropriate choice of parameters, the
  integral of the autocorrelation function of the model and the truth
  is identical, presumably leading to a more similar climatological
  response.}
\label{changeAutoCorr:fig}
\end{figure}

The fluctuation-dissipation theorem (e.g. \citealp{Marconi08})
guarantees, under very general assumptions, that the sensitivity to a
small forcing is equal to a time integral of the correlation of some
function of the prognostic variables. Thus fluctuation-dissipation
theorem provides a method by which we can relate the response of a
complex system to variables that we can measure. In the case of a
single variable linear stochastic system the sensitivity is simply the
time integration of its autocorrelation. For application to a general
circulation model see \cite{Gritsun07}. Unfortunately the response
according to its autocorrelation function when assuming a linear
stochastic model may not be accurate \citep{Cooper14} and the
true response of a chaotic non-linear system to forcing is not
straightforward to evaluate using the fluctuation-dissipation theorem
(\citealp{cooperetal2011}, \citealp{Cooper13}). Despite those
limitations, one can reasonably assume that a model will respond in a
more similar manner to the truth it represents, the more the
autocorrelation functions of the respective systems are alike.  We
therefore require a parameterisation that pushes the model to have an
autocorrelation function to be close to truth. Given that a stochastic
term is characterised by an amplitude and a time scale, it can
therefore be used to adjust both the variance and lag-covariance of a
model towards the true climatological values. In our case the
stochastic term is represented in a discretized model by a system of
linear stochastic ordinary differential equations
\begin{equation}
d \bm{\xi} = \mathbf{B}\bm{\xi} dt + \mathbf{V} \sqrt{\mathbf{D}} d\mathbf{w}.
\label{stochasticTerm:eqn}
\end{equation}
Here $\mathbf{B}$ is a constant matrix, restricted to have all
negative eigenvalues so that the system is stable, $\bm{\xi}$ is a
vector of model grid points, each element of which is to be added to
the right hand side of the discretized governing equations,
$\mathbf{w}$ denotes a vector of independent Wiener or white noise
processes with unit variance. The covariance of the stochastic term is
set by the matrix $\mathbf{Q}$ with the matrix of eigenvectors of
$\mathbf{Q}$ denoted by $\mathbf{V}$ and diagonal matrix of
eigenvalues by $\mathbf{D}$. Changing the values of $\mathbf{B}$ and
$\mathbf{Q}$ allows us to set the integral of the autocorrelation to
approximately the true value, and hence improve the accuracy of the
model's response to a small forcing, see Figure
\ref{changeAutoCorr:fig}.


\subsection{A simple parameterisation scheme}

In summary we propose adding a spatially-dependent but constant in
time forcing term (section \ref{mean:sect}) and a linear stochastic
term (sections \ref{variance:sect} and \ref{ForcedRespIntro:sect}) to
the equations governing a turbulent fluid. The aim being to reproduce
some of the statistics of a low viscosity fluid with a high viscosity
numerical integration. In the present work, we focus on finding the
idealised mean sub-grid barotropic eddy forcing terms in an ocean
double gyre configuration. The forcing is limited to the momentum
sub-grid forcing in a shallow-water model, unlike the study of
\citet{ferreiraetal2005} which tackled the eddy buoyancy forcing in a
primitive equation model.  The approach is similar to a standard
linear relaxation where prognostic variables are linearly forced
towards some basic state (e.g. in the configuration described in
\citealp{Held94}). However, in our case the variable being relaxed, in
a stochastic sense, is independent of the unparameterised prognostic
variables.  \cite{berloff2005} applies a similar scheme with
parameters diagnosed from single high and low resolution integrations
to a quasi-geostrophic system.  The novel approach used in the present
work is to optimise the values of these parameters and to develop a
method that is applicable to complex models involving more than one
prognostic variable.  We optimise a low resolution model in such a way
as to improve its characteristics when compared with the truth as
defined by observations or a high resolution integration. Our scheme
involves empirically derived parameters and therefore we refer to it
as a parameterisation scheme. However given the number of parameters,
three for each grid cell, that the scheme captures the bulk effect of
unspecified sub-grid scale processes as opposed to mimicking specific
processes and the fact that no dependence on the large-scale
flow or external parameters is found, some readers may not consider it
to be a parameterisation scheme in the conventional sense. Calling it a
bias correction scheme may be more appropriate, albeit in a free running model.

Our choice of test model is primarily motivated by the more general,
and to the authors knowledge unsolved, optimisation problem for the
long time statistics of a high dimensional chaotic system.  We hope
that our particular configuration, described in section
\ref{model:sect}, contains no special features that make the
optimisation problem solvable in only this case. However further
investigations would be required to check the applicability in other
physical configurations such as a periodic channel model and moving
from a two dimensional, to a three dimensional fluid.

Integrations performed using a shallow water model are detailed in
Section \ref{model:sect} and the optimisation method in Section
\ref{method:sect}.  In Section \ref{results:sect}, we compare the high
and low resolution integrations without parameterisation with the
optimised low resolution system and investigate the effect of the
parametrisation
scheme. 

\section{The truth and model integrations}
\label{model:sect}

Our choice of an idealised ocean model is motivated by the ocean
double gyre setup of \cite{berloff2005} although rather than integrate
the quasi-geostrophic equation, we are integrating the shallow water
equations with a linear free surface.  To reduce the computational
cost, our system is barotropic and therefore omits the baroclinic
modes and their interaction with the mean flow.  Although the
resulting setup is idealised it is sufficient to illustrate our
parameterisation scheme.

\begin{figure}
\begin{center}
\includegraphics[width=0.49\textwidth]{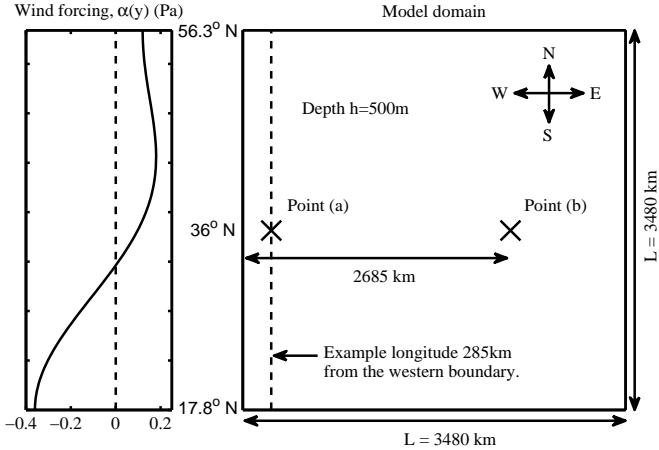}
\end{center}
\caption{The wind forcing profile and integration domain, see
  text. The autocorrelation of the sea surface height, $\eta$ at
  points (a), (b) is presented in Figure \ref{sshCorr:fig}. The
  climatological mean sea surface height and eddy kinetic energy at
  the longitude indicated is presented in Figures
  \ref{climMeanSSH:fig} and \ref{climEKE:fig} respectively.}
\label{domain:fig}
\end{figure}

The integration domain (Figure \ref{domain:fig}) is a 3840 km by 3840
km rectangular box located on a mid-latitude beta plane with flat
bottom. The flow is forced by a (time) constant surface wind which
varies latitudinally (see Figure \ref{domain:fig}) but is uniform in
longitude. It is dissipated by a viscous term whose magnitude
depends upon the resolution of the model grid. The ocean floor and
horizontal boundaries are free slip. Berloff's domain was uniformly 3
km deep, however for the magnitude of forcing chosen, a 3 km deep
barotropic system exhibits extremely long timescales at low
resolution. Such long timescales makes the experiments rather
expensive. A simple solution to reduce the cost is by increasing the
turbulent nature of the flow that in turn reduces the correlation
times present and makes estimation of the autocorrelation
computationally feasible. The system is made more turbulent by
reducing the ocean depth to a constant 500 m and increasing the wind
forcing by a factor of three. Unfortunately this more chaotic system
might resemble even less a realistic ocean gyre but nevertheless
should not affect the main conclusions.

The equations governing the time evolution of the prognostic
variables, zonal velocity $u$ ($x$ direction), meridional velocity $v$
($y$ direction) and sea surface height $\eta$ are
\begin{align}
\frac{\partial u}{\partial t} + u\frac{\partial u}{\partial x} + v\frac{\partial u}{\partial y} &- fv + g \frac{\partial \eta}{\partial x} - \kappa \nabla^2 u \notag \\
&= \frac{1}{\rho_0 h} \left( \alpha(y) + F_u(x,y) + \xi_u(x,y,t) \right), \label{dudt:eqn} \\
\frac{\partial v}{\partial t} + u\frac{\partial v}{\partial x} + v\frac{\partial v}{\partial y} &+ fu + g \frac{\partial \eta}{\partial y} - \kappa \nabla^2 v \notag \\
&= \frac{1}{\rho_0 h} \left( F_v(x,y) + \xi_v(x,y,t) \right), \label{dvdt:eqn} \\
\frac{\partial \eta}{\partial t} + h \left( \frac{\partial u}{\partial x} + \frac{\partial v}{\partial y} \right)
&= 0,
\label{dEtaDt:eqn}
\end{align}
where the acceleration due to gravity $g=9.81$ m~s$^{-2}$, density of water $\rho_0=999.8$ kg m$^{-3}$, depth $h=500$ m and the Coriolis parameter $f=f_0+\beta y$ with $\beta=2 \times 10^{-11}$ m$^{-1}$s$^{-1}$ and $f_0=4.46 \times 10^{-5}$ s$^{-1}$. $\alpha(y)$ represents a zonal wind forcing of the form
\begin{equation}
\alpha(y) = \alpha_0 \left[ \cos \left( \frac{2\pi\left( y-L/2 \right)}{L} \right) + 2 \sin \left( \frac{\pi \left( y-L/2 \right)}{L} \right)\right],
\end{equation}
where the domain width in each direction $L=3840$ km with $0 \leq y
\leq L$, $0 \leq x \leq L$ corresponding to a domain bounded between
latitudes $17.8^{\circ}$ N and $56.3^{\circ}$ N. The constant
$\alpha_0=0.12$ Pa. The model equations (\ref{dudt:eqn}),
(\ref{dvdt:eqn}) and (\ref{dEtaDt:eqn}) are discretized onto a uniform
Cartesian Arakawa C-grid \citep{Arakawa77}, initialised to zero or
with low amplitude random noise, and integrated using the MITgcm
\citep{Marshall97}. 

In the parameterised low resolution system the constant in time but
spatially varying forcing $F_u(x,y)$ and $F_v(x,y)$, represented by the
vectors $\mathbf{f}_u$ and $\mathbf{f}_v$, are found by
optimisation. $\xi_u(x,y,t)$ and $\xi_v(x,y,t)$ are stochastic terms,
represented by $\bm{\xi}_u$ and $\bm{\xi}_v$. $\bm{\xi}_u$ and
$\bm{\xi}_v$ would, in a fairly general case, be governed by equation
(\ref{stochasticTerm:eqn}) with an appropriate choice of
$\mathbf{B}_u$, $\mathbf{B}_v$, $\mathbf{Q}_u$ and $\mathbf{Q}_v$
found via the optimisation. We consider the simpler case where all
elements of $\bm{\xi}_u$ and $\bm{\xi}_v$ are independent of each
other (uncorrelated in space) and the equations for their time
evolution reduce to
\begin{equation}
d {\xi_u}_i = {b_u}_i {\xi_u}_i dt + \sqrt{{q_u}_i} {dw_u}_i
\label{stochasticU:eqn}
\end{equation}
and
\begin{equation}
d {\xi_v}_i = {b_v}_i {\xi_v}_i dt + \sqrt{{q_v}_i} {dw_v}_i
\label{stochasticV:eqn}
\end{equation}
for $i=1 \dots d$ where $d$ is the number of grid cells in the
integration and ${w_u}_i$ and ${w_v}_i$ are white noise processes with
unit variance. In (\ref{stochasticU:eqn}) and (\ref{stochasticV:eqn})
the variables are the elements of the vector quantities
$\bm{\xi}_u=\left({\xi_u}_1,{\xi_u}_2,\cdots \right)$,
$\bm{\xi}_v=\left({\xi_v}_1,{\xi_v}_2,\cdots \right)$ and the
constants $\mathbf{b}_u=\left({b_u}_1,{b_u}_2,\cdots \right)$,
$\mathbf{b}_v=\left({b_v}_1,{b_v}_2,\cdots \right)$,
$\mathbf{q}_u=\left({q_u}_1,{q_u}_2,\cdots \right)$ and
$\mathbf{q}_v=\left({q_v}_1,{q_v}_2,\cdots \right)$ represent the
diagonals of $\mathbf{B}_u$, $\mathbf{B}_v$, $\mathbf{Q}_u$ and
$\mathbf{Q}_v$ respectively. Test integrations show that the addition
of $\xi_u(x,y,t)$ and $\xi_v(x,y,t)$ in the coarse resolution set-up,
resulting from equations (\ref{stochasticU:eqn}) and
(\ref{stochasticV:eqn}), do not cause the model to have a systematic
long term drift.

The high resolution ``truth'' has a grid spacing of $\Delta x = \Delta
y = 7.5$ km corresponding to 512 by 512 grid cells, a viscosity of
$\kappa=10$ m$^2$s$^{-1}$ and $f_u=f_v=\xi_u=\xi_v=0$. The truth is
integrated for $10^4$ days after discarding a spin up of $10^3$
days. $10^4$ days is chosen so as to provide estimates of the
climatological mean, variance and lag-covariance with sufficient
accuracy. A grid spacing of $\Delta x = \Delta y = 30$ km
corresponding to 128 by 128 grid cells, a viscosity of $\kappa=470.23$
m$^2$s$^{-1}$ integrated for the same time is used to represent a low
resolution ``model'' of the ``truth''.  The viscosities are chosen for
stability to yield a Munk layer width $M_w$ along the western boundary
of at least three grid cells, $M_w=\pi \left( \kappa / \beta
\right)^{1/3} > 3 \Delta x$.  Throughout this paper we refer to high
7.5 km resolution integrations as the truth and low 30 km resolution
integrations as the model.  The initial model without additional
constant or stochastic forcing is referred to as the unparameterised
model and the optimised model with the additional constant and
stochastic forcing is referred to as the parameterised model. To
compare the model with the truth, linearly interpolated values of the
truth integration at the locations of the low resolution model grid
cell variables are used.

\section{The optimisation algorithm}
\label{method:sect}

For the discretized $u$ and $v$ fields, dropping the subscript for
simplicity, the parameters $\mathbf{f}$ (representing a
time-independent spatially varying forcing), $\mathbf{b}$ and
$\mathbf{q}$ (governing the respective timescale and amplitude of the
stochastic process), are estimated using an iterative process. These
vectors are initialised to zero, $\mathbf{f}^0=0$, $\mathbf{b}^0=0$
and $\mathbf{q}^0=0$ where the superscript indicates an iteration
number $n$. For each iteration, the low resolution model is
integrated. After an initial spin up period, the model climatological
mean of $u$ or $v$ at each grid point, denoted
$\mathbf{m}_{\text{model}}^n$, is estimated by integrating over a
sufficiently long time (in our case $10^4$ days). The model
climatological mean is compared with the true climatological mean
vector $\mathbf{m}_{\text{truth}}$ and $\mathbf{f}$ is updated as
follows
\begin{equation}
\mathbf{f}^{n+1}=\mathbf{f}^n + \left( \mathbf{m}_{\text{truth}} - \mathbf{m}_{\text{model}}^n \right) \delta f,
\end{equation}
where $\delta f$ is a suitably small positive constant. A similar
procedure is followed for updating $\mathbf{q}$
\begin{equation}
\mathbf{q}^{n+1}=\mathbf{q}^n + \left( \bm{\sigma}_{\text{truth}}^2 - \left(\bm{\sigma}_{\text{model}}^n\right)^2 \right) \delta q,
\end{equation}
where $\bm{\sigma}_{\text{model}}^n$ is the standard deviation of the
low resolution parameterised model $u$ or $v$ at each grid point
measured at the $n$'th iteration, $\bm{\sigma}_{\text{truth}}$ is the
standard deviation of the true system and $\delta q$ is again a
suitably small positive constant. The minimum value that $\mathbf{q}$
can obtain is clipped at zero, corresponding to no stochastic forcing
at the relevant grid point.

\begin{figure}
\includegraphics[width=0.49\textwidth]{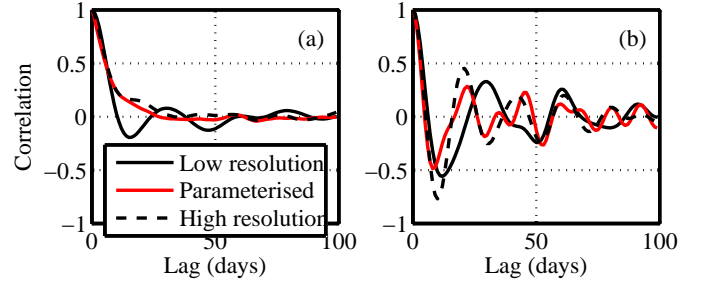}
\caption{The autocorrelation of the sea surface
  height of each low resolution model and the high resolution truth at $36^{\circ}$ latitude, (a) 285 km and (b) 2685 km from the
  western boundary, see Figure \ref{domain:fig}. Point (a) is right in
  the centre of the most turbulent part of the domain where the
  largest value of the mean eddy kinetic energy is found (compare with
  \cite{berloff2005}, Figure 4(a)). Point (b) is in a region dominated
  by wave activity, the autocorrelation function is more oscillatory
  with the oscillations taking longer to decay (compare with
  \cite{berloff2005}, Figure 4(b)).}
\label{sshCorr:fig}
\end{figure}

As mentioned in section \ref{ForcedRespIntro:sect}, we wish to set
$\mathbf{b}$ such that the integral of the lag-covariance (or
equivalently the integral of the autocorrelation function), of the
model and the truth are the same. From data, estimates of the integral
of the lag-covariance are not as accurate as estimates of the
lag-covariance at a particular lag \citep{Cooper14}. Therefore we aim
to choose a single lag $\tau$ and optimise $\mathbf{b}$ so that the
model and the truth have the same autocorrelation at this lag. The
idea is that if the model and the truth have the same autocorrelation
at this lag, they also approximately have the same integral of the
autocorrelation. The autocorrelation of the stochastic term is an
exponential decay. It turns out that choosing a lag that is too small
leads to an overestimate of the integrated lag-covariance because the
initial decay in the truth autocorrelation function is slower than
exponential \citep{DelSole00}. On the other hand the uncertainty in
the autocorrelation as a percentage increases with lag, so a lag that
is too large leads to a large uncertainty in the decorrelation
time. By plotting the autocorrelation of a selection of the high
resolution variables against lag (as in Figure \ref{sshCorr:fig})
over exponentials with various decay constants, we
estimate that the two curves meeting at a lag $\tau=5$ days gives a
reasonable, but imperfect, exponential approximation to the autocorrelation.
This is
similar to the value that would be obtained by \cite{berloff2005}, see
their Figure 4.
Figure \ref{sshCorr:fig} demonstrates
that the autocorrelation varies across the domain, so as an
alternative to our choice of $\tau = 5$ days, it might be more
reasonable to assume that the appropriate $\tau$ to use is found when
the autocorrelation first reaches 0.6 or some other reasonable
value.

It can be shown, see \ref{linearParam:sect}, that for linear systems,
$-1/b_i$ is approximately linearly related to the lag covariance
introduced to the system, where $b_i$ is element $i$ of
$\mathbf{b}$. Therefore the procedure for updating $\mathbf{b}$ is
given by
\begin{equation}
\frac{-1}{b^{n+1}_i}=\frac{-1}{b^n_i} + \left( c_{i,\text{truth}} (\tau) - c_{i,\text{model}}^n(\tau) \right) \delta b
\label{iterateB:eqn}
\end{equation}
for $i=1 \dots d$. Here $c_{i,\text{truth}}(\tau)$ and
$c_{i,\text{model}}^n(\tau)$ represent the lag $\tau$ covariance,
independent at each grid point, for the respective truth and $n$'th
model iteration. $\delta b$ once again is a sufficiently small
positive constant. $-1/b_i$ is constrained to be always greater than
600 seconds. Each element of $\mathbf{b}$ is therefore always negative
and the stochastic systems (\ref{stochasticU:eqn}) and
(\ref{stochasticV:eqn}) are guaranteed to be stable.

The constants $\delta f=0.1$ kg m$^{-2}$ s$^{-1}$, $\delta q=0.1$
kg$^2$ m$^{-4}$ s$^{-3}$, and $\delta b=5.0$ m$^{-2}$ s$^3$, were
found by trial and error to lead to convergence. If they are too
small, the algorithm is too slow to converge and if they are too large
this Euler type method is unstable. The model's initial condition at
each iteration of the optimisation is the final state taken from the
end of the previous iteration. Using this method it was found that it
was only necessary to discard a spin up of 500 days at the start of
each iteration. When the difference between the truth and model
climatological state at a grid point becomes smaller than the
uncertainty in the mean, variance and 5 day lag covariance, further
optimisation is not possible. The integration length governs this
uncertainty. Longer (and more computationally expensive) integrations
than the $10^4$ days used here is found to lead to faster convergence
per iteration and the optimisation converges to a more accurate state.

We do not make use of additional assumptions, such as divergence free
forcing. Although it would be interesting to see if such assumptions
can improve our results, our aim here is generality, and it is not
clear that they would be appropriate in all cases. Removing the
requirement to restrict optimisations based on assumed conservation
laws or other physical properties, means that understanding the
perhaps unknown physics of the complex system is not necessary for
optimisation.  If optimisation is successful, then the optimised
system will obey and give insight into the appropriate physical
laws. However further investigations would be required to check the
applicability in other physical configurations such as a periodic
channel model and moving from a two dimensional, to a three
dimensional fluid.

\section{Results}
\label{results:sect}

\begin{figure*}
\begin{center}
\includegraphics[width=\textwidth]{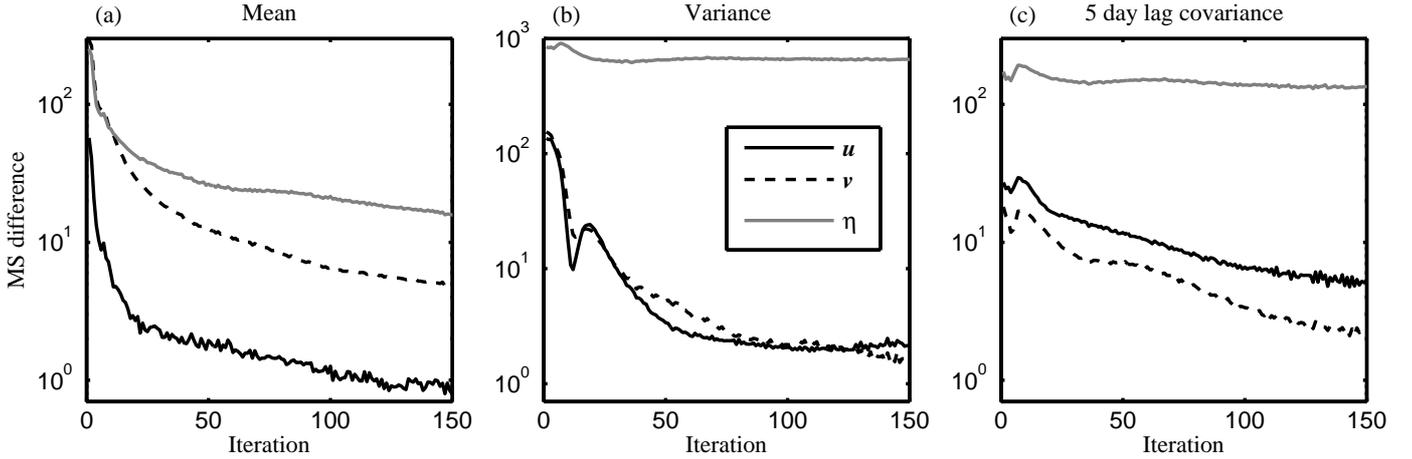}
\end{center}
\caption{The mean squared (MS) difference between the parameterised
  model and high resolution truth climatological (a) mean, (b)
  variance and (c) 5 day lag covariance for the zonal velocity $u$
  (solid), meridional velocity $v$ (dashed) and sea surface height
  $\eta$ (grey). Note the logarithmic $y$ axis. The unparameterised
  model represents the first iteration.}
\label{L2NormVsIter:fig}
\end{figure*}

In our barotropic double gyre configuration, the parameters control the sub-grid eddy momentum forcing. The optimisation algorithm finds a slightly improved set of parameters at each iteration.
With the first point corresponding to
the unparameterised model, the mean squared differences between the
high resolution and parameterised climatological mean, variance and 5
day lag covariance are plotted with a logarithmic $y$ axis in Figure
\ref{L2NormVsIter:fig}. Using this metric, the difference between the
parameterised and true climatology has been substantially reduced when
compared with the unparameterised low resolution system. After 150
iterations the mean $u$, $v$ and $\eta$ fields are continuing to
improve; however the variance of all three fields seems to have
reached a plateau. The 5 day lag covariance of the $u$ and $v$ fields
are still improving while for $\eta$ it seems to have also
plateaued. The minimal improvement in $\eta$,
which has no sub-grid forcing applied to it's governing equation,
is examined in more detail in section \ref{climVar:sect}.

\subsection{The climatological mean}

\begin{figure*}
\includegraphics[width=\textwidth]{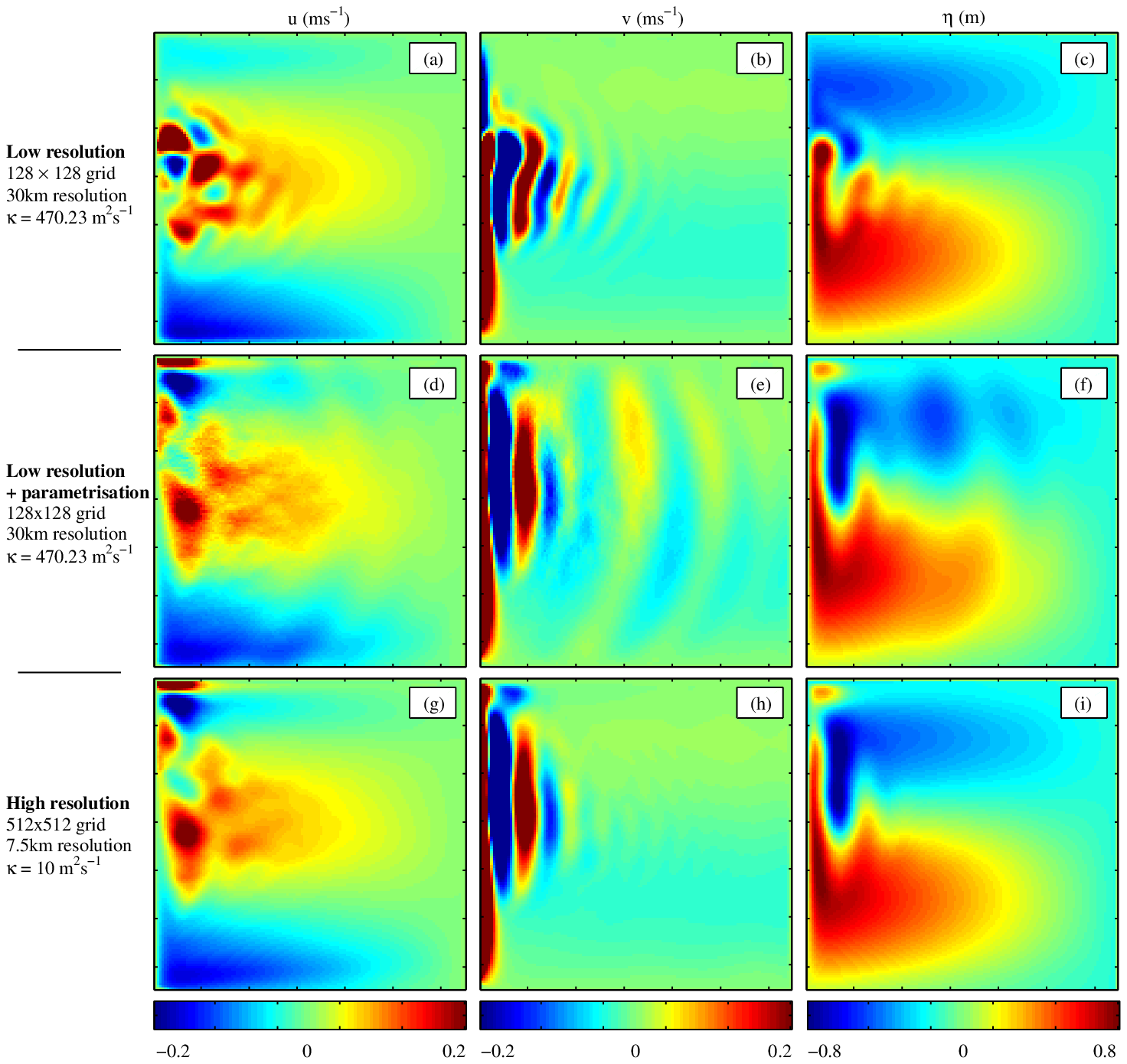}
\caption{The climatological mean zonal velocity $u$ (left), meridional
  velocity $v$ (middle) and sea surface height $\eta$ (right) for the
  low resolution (top), low resolution parameterised at iteration 150
  (middle), and high resolution model (bottom). $\kappa$ is the
  viscosity parameter.}
\label{climMean:fig}
\end{figure*}

\begin{figure}
\includegraphics[width=0.49\textwidth]{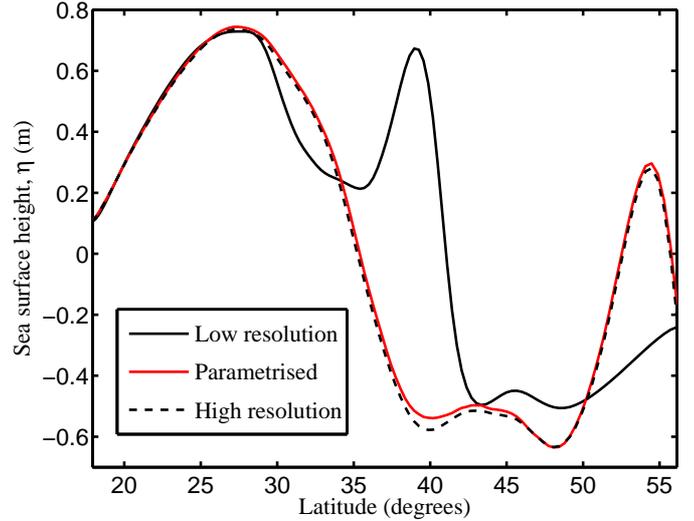}
\caption{The climatological mean sea surface height $\eta$ at a single
  longitude 285 km from the western boundary, see Figure
  \ref{domain:fig}.}
\label{climMeanSSH:fig}
\end{figure}

After 150 iterations, Figure \ref{climMean:fig} shows that the
climatological mean state of the low resolution parameterised model
(panels (d), (e) and (f)) is significantly closer to the high
resolution mean (panels (g), (h) and (i)) when compared with the low
resolution model without parameterisation (panels (a), (b) and
(c)). The large values of $u$, $v$ and $\eta$ near the western
boundary are correctly reproduced and the magnitude and shape
of the zonal barotropic jet has been improved.
Taking a single longitude of the climatological mean sea
surface height 285 km from the western boundary, Figure
\ref{climMeanSSH:fig} demonstrates that the parameterised model
adequately mimics the truth. For example at $40^{\circ}$ latitude, 285
km from the western boundary, $\eta$ is 51.3 cm, -53.9 cm and -57.8 cm
for the unparameterised, parameterised and truth integrations
respectively. Note the change in sign. The relatively small
  difference between the high resolution and parameterised
  integrations, spread unevenly over the whole domain, stem from the
  fact that the system is chaotic. Estimation of the climatological
  mean of a chaotic system is subject to some error, proportional to
  $1/\sqrt{t_{\text{max}}}$ where $t_{\text{max}}$ is the integration
  length. Therefore increasing $t_{\text{max}}$ reduces this
  error. Constant forcing of the $u$ and $v$ fields leads to
  optimisation of the $\eta$ field because of approximate geostrophic
  balance. Additional forcing of $\eta$ by adding a spatially
  dependent term, $F_{\eta}(x,y)$, to (\ref{dEtaDt:eqn}) introduces
  additional sources and sinks of mass, but may also lead to small
  improvements because neither geostrophic balance or the numerical
  scheme used is exact.

\subsection{The climatological variance}
\label{climVar:sect}

\begin{figure*}
\includegraphics[width=\textwidth]{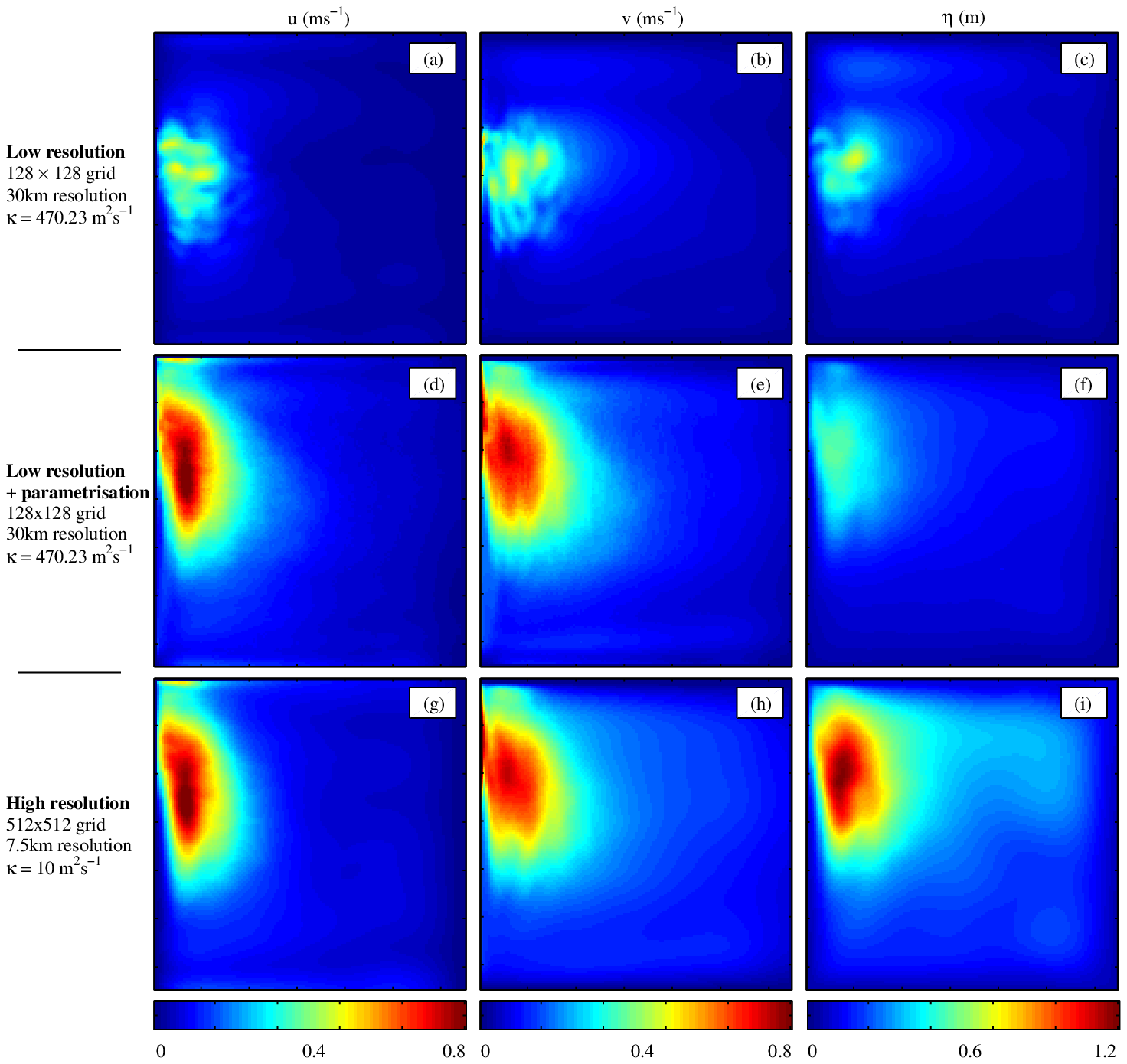}
\caption{The climatological standard deviation of the zonal velocity
  $u$ (left), meridional velocity $v$ (middle) and sea surface height
  $\eta$ (right) for the low resolution (top), low resolution
  parameterised at iteration 150 (middle), and high resolution model
  (bottom). $\kappa$ is the viscosity parameter.}
\label{climVar:fig}
\end{figure*}

\begin{figure*}
\includegraphics[width=\textwidth]{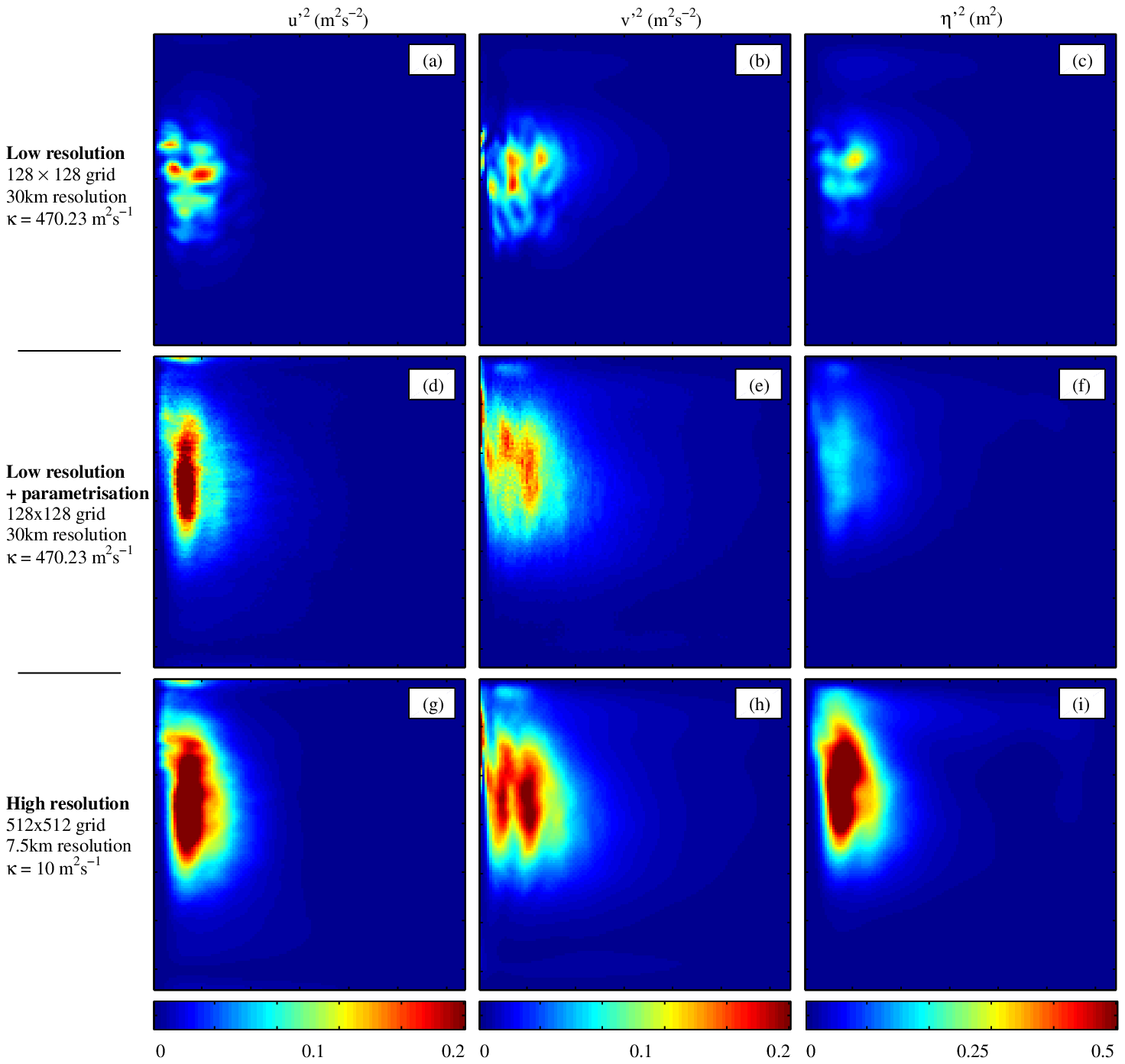}
\caption{The climatological 5 day lag covariance of the zonal velocity
  $u$ (left), meridional velocity $v$ (middle) and sea surface height
  $\eta$ (right) for the low resolution (top), low resolution
  parameterised at iteration 150 (middle), and high resolution model
  (bottom). $\kappa$ is the viscosity parameter.}
\label{climLag5Var:fig}
\end{figure*}

Figures \ref{climVar:fig} and \ref{climLag5Var:fig} show that the low
resolution unparameterised model (panels (a), (b) and (c)) has a much
lower climatological variance and 5 day lag covariance than the true
high resolution model (panels (g), (h) and (i)). The optimisation has
succeeded in increasing the variance and 5 day lag covariance of the
low resolution parameterised model to be much closer to that of the
high resolution system for the $u$ and $v$ fields (panels (d) and
(e)). The 5 day lag autocorrelation, Figure \ref{sshCorr:fig}, indicates
that the long persistence of the low
resolution model, relative to the high resolution model, has been
reduced. Decomposing the velocities into their mean and varying
components, $u=\bar{u}+u'$ and $v=\bar{v}+v'$, the eddy kinetic
energy, defined as $\frac{1}{2} \Delta x \, \Delta y \, h \, \rho_0 (
u'^2+v'^2 )$, is proportional to the variance in the $u$ and $v$
fields. Taking the
climatological eddy kinetic energy at a single longitude, 285 km from
the western boundary, Figure \ref{climVar:fig} demonstrates that the
parameterised model is again a good representation of the truth. For
example at $40^{\circ}$ latitude, 285 km from the western boundary,
within a 30 km grid cell it is $0.6 0\times 10^{14}$ J, $2.32 \times
10^{14}$ J and $2.34 \times 10^{14}$ J for the unparameterised,
parameterised and truth integrations respectively. Both the variance
and 5 day lag covariance in the $\eta$ field is not so well reproduced
(compare panels (f) and (i)).

$\eta$ is well approximated by the two dimensional stream
  function, $\psi$, defined by $u=-\partial \psi/\partial y$ or
  $v=\partial \psi/\partial x$. $\psi$, and hence $\eta$, can
  therefore be approximated as a spatial integral over $u$ or $v$. The
  noise terms $\xi_u$ and $\xi_v$ are uncorrelated in space, so their
  integral over the domain tends to be small. A large positive $\xi_u$
  at one latitude is likely to be cancelled out by negative values of
  $\xi_u$ at other latitudes. Thus the contribution of $\xi_u$ and
  $\xi_v$ to the variability of $\eta$ turns out to be relatively
  small. To get the correct variability of the $\eta$ field a
  stochastic term needs to be added to the right hand side of the
  equation for $\eta$, (\ref{dEtaDt:eqn}). Alternatively, including
  spatial correlation (represented by the off diagonals of
  $\mathbf{Q}$), when integrating equations (\ref{stochasticU:eqn})
  and (\ref{stochasticV:eqn}) would increase the variance of the
  spatially integrated $\xi_u$ and $\xi_v$. Therefore the variance in
  $\eta$ would also increase. Either of these options requires more
  investigation and are beyond our current scope.


\begin{figure}
\includegraphics[width=0.49\textwidth]{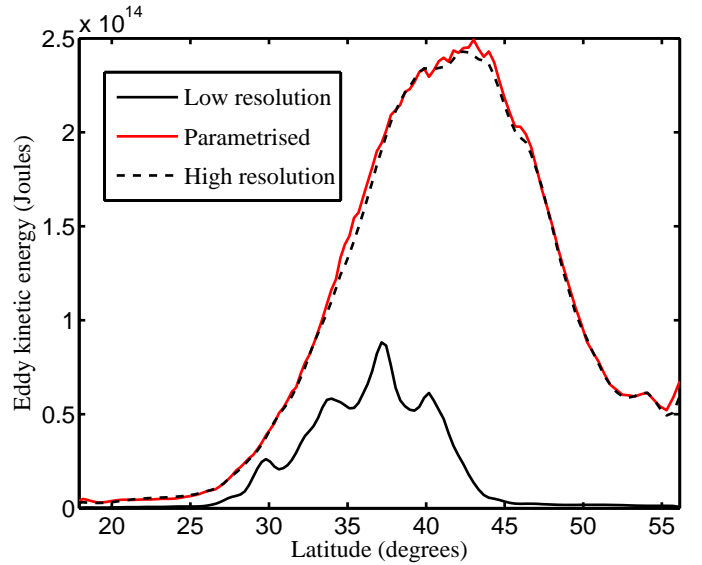}
\caption{The climatological eddy kinetic energy within each 30km grid
  cell at a single longitude 285 km from the western boundary, see
  Figure \ref{domain:fig}.}
\label{climEKE:fig}
\end{figure}

\subsection{The constant forcing parameters}
\label{constForce:sect}

The parameterisation terms added to the equations for $u$ and $v$
found by the optimisation at iteration 150 are described in Figure
\ref{paramVals:fig}. The constant sub-grid momentum forcing term
necessary to maintain the recirculating gyre, $F_u$, applied to the
equation for $u$ (panel (a)) is dominated by a large positive value,
denoting forcing fluid from west to east, in the north-west corner of
the domain. There are however patches of forcing with amplitude
changing rapidly over small length scales, close to that of the grid
scale, in parts of the turbulent region.

This may be due either to slow convergence of the optimisation
algorithm (perhaps the true forcing is smooth) or due to uncertainty
in the climatological mean stemming from the finite integration
length. Perhaps the small scales in the forcing structure are
necessary as suggested in other studies (e.g. \citealp{kraichnan1976},
\citealp{zidikherietal2010}, \citealp{Mana14} and references
therein). This was tested by smoothing the iteration 150 $F_u$ and
$F_v$ fields using a nearest neighbour average for each grid
cell. This smoothing reduced the quality of the climatological mean
and demonstrates that at least some small-scale features are needed
for an optimal solution.

The constant forcing of the $v$ field, $F_v$ (Figure
\ref{paramVals:fig}, panel (d)), is very strong (northwards) for a
single grid box alongside the western boundary. For example at
$40^{\text{o}}$ latitude $F_v=1.25$ Pa. It reaches a maximum of $1.32$
Pa at $36.6^{\text{o}}$ latitude which compares with a maximum, over
the remainder of the domain, of $0.62$ Pa at a point in the red
(colour saturated) region close to, but not always alongside, the
western boundary. In the main body of the domain there are alternate
bands of northward and southward forcing approximately aligned with
the western boundary. A possible explanation for the two strongest
bands closest to the western boundary is that they extend the
meridional flow patterns, and hence the flow boundary separation
point, northwards. In the low resolution model without
parameterisation these flow patterns stop half way along the domain,
compare with Figure \ref{climMean:fig} panel (b).  Figure
\ref{climMean:fig} also shows some differences between the low
resolution parameterised and the high resolution models far from the
western boundary (compare panels (e) and (h)). Perhaps these
differences are due to a finite integration time giving a
climatological mean that is not perfectly resolved. The shape of these
differences reflects the linear Rossby waves present in this region
and the optimisation algorithm may be trying to correct for these
differences but inadvertently magnifying them. On the other hand, the
forcing elsewhere may lead to a correction far from the western
boundary that the optimisation is in turn trying to correct. Longer
integrations or more iterations may resolve this issue.

Now that the values of $F_u$ and $F_v$ have been found, it is interesting to consider which physical processes set the pattern of the forcing.  In the barotropic double gyre experiment, we expect the non-linear (Reynolds stresses) and viscous terms to dictate the mean sub-grid eddy forcing. Examination of these terms in the truth and low resolution experiment indicate that forcing of the form
\begin{equation}
F_u \approx \gamma_1 \nabla^2 \overline{u_{\text{T}}}
\qquad and \qquad
F_v \approx \gamma_1 \nabla^2 \overline{v_{\text{T}}},
\label{guessConstForce:eqn}
\end{equation}
might be a reasonable approximation. Here $\gamma_1$ is a constant, the over bar indicates the time mean and the subscript T indicates that data from the truth integration is used. Forcing the model with these terms, (with $\gamma_1 = 4 \times 10^9$ kg s$^{-1}$ chosen to approximately match the amplitude of the optimised forcing shown in Figure  \ref{paramVals:fig}), leads to the separation point of the jet being further north and encouraging changes to the structure of the eastwards jet, see Figure \ref{guessForcing:fig}. However the strength and pattern is significantly different from the high resolution truth suggesting that if this form of forcing plays a role, non-linear feedbacks are important.


\subsection{The stochastic forcing parameters}
\label{stochForce:sect}

\begin{figure*}
\includegraphics[width=\textwidth]{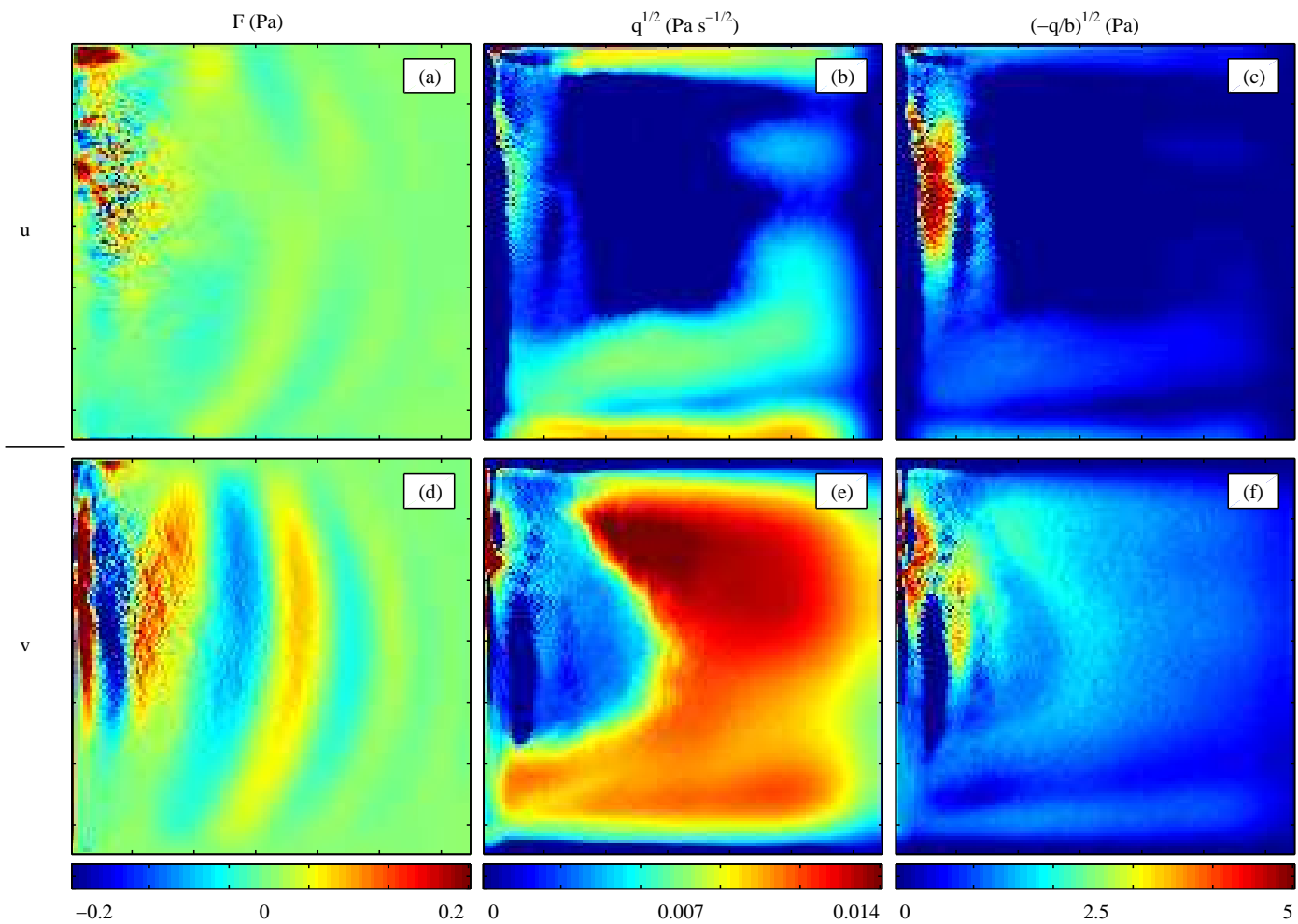}
\caption{Iteration 150 of the optimised values for the zonal $u$ (top)
  and meridional $v$ (bottom) velocity forcing parameters $\mathbf{f}$
  (left) and stochastic parameters $\mathbf{q}$ (middle) and
  $\left(-\mathbf{q}/\mathbf{b}\right)^{1/2}$ (right) where the
  division and square root is taken separately for each vector
  element. The quantity $\left(-\mathbf{q}/\mathbf{b}\right)^{1/2}$
  has Pascals as its units and is a measure of the strength of the
  stochastic forcing.}
\label{paramVals:fig}
\end{figure*}

The variance $\mathbf{q}_u$ of the noise in the stochastic system
governing $\bm{\xi}_u$ is relatively large at a few grid points in the
north-west corner of the domain and along the northern and southern
boundaries (Figure \ref{paramVals:fig}, panel (b)). There is also
noise in the vicinity of the high resolution eddy activity, but there
are regions in the centre of the domain and all along the western
boundary where variance of $\bm{\xi}_u$ is zero. $\mathbf{q}_v$ on the
other hand (Figure \ref{paramVals:fig}, panel (e)) is large close to
the northern part of the western boundary, small or zero along the
northern and southern boundaries, small in the region of eddy activity
and large in the region of linear wave activity. Also there is an
oval region at the centre of eddy activity where $\mathbf{q}_v$ is
zero.

A measure of the effective forcing of the stochastic term is given by
$\sqrt{-\mathbf{q}_u/\mathbf{b}_u}$ and
$\sqrt{-\mathbf{q}_v/\mathbf{b}_v}$ with units of Pascals and the
division denotes the element wise division of each element of
$\mathbf{q}$ by the corresponding element of $\mathbf{b}$. The square
root is also taken element wise. Thus in addition to $\mathbf{q}_u$
and $\mathbf{q}_v$, the time scale of $\bm{\xi}_u$ and $\bm{\xi}_v$
determined by $\mathbf{b}_u$ and $\mathbf{b}_v$ needs to be taken into
account. Panels (c) and (f) of Figure \ref{paramVals:fig} show that
the largest stochastic forcing is located in the region of high
resolution eddy activity. In this region, the amplitude of the
stochastic term is small, but the time scale, of around 10 to 12 days,
is relatively long when compared with the time scale in the rest of
the domain, $\sim 5$ hours. In \cite{berloff2005} stochastic forcing
was applied using a first and second order auto-regressive
process. For various experiments their first order process was given a
spatially uniform time scale between 3.3 and 30 days.

To further try to disentangle the overlapping effects of the four
additional forcing terms $F_u$, $F_v$, $\xi_u$ and $\xi_v$, a set of
experiments was performed where a subset of these terms was set to
zero and the optimisation was performed again. These experiments,
summarised in table \ref{optimiseWhat:table}, show that using the full
set of four terms leads to the most optimal climatological means and
variances. With this experimental configuration the constant forcing
terms $F_u$ and $F_v$ are largely responsible for correcting the
climatological mean and the stochastic terms, $\xi_u$ and $\xi_v$, are
mainly responsible for correcting the climatological variance and 5
day lag covariance. In addition, it is crucial that both the amplitude
and time scale of the stochastic term is allowed to vary spatially.

\begin{figure*}
\includegraphics[width=\textwidth]{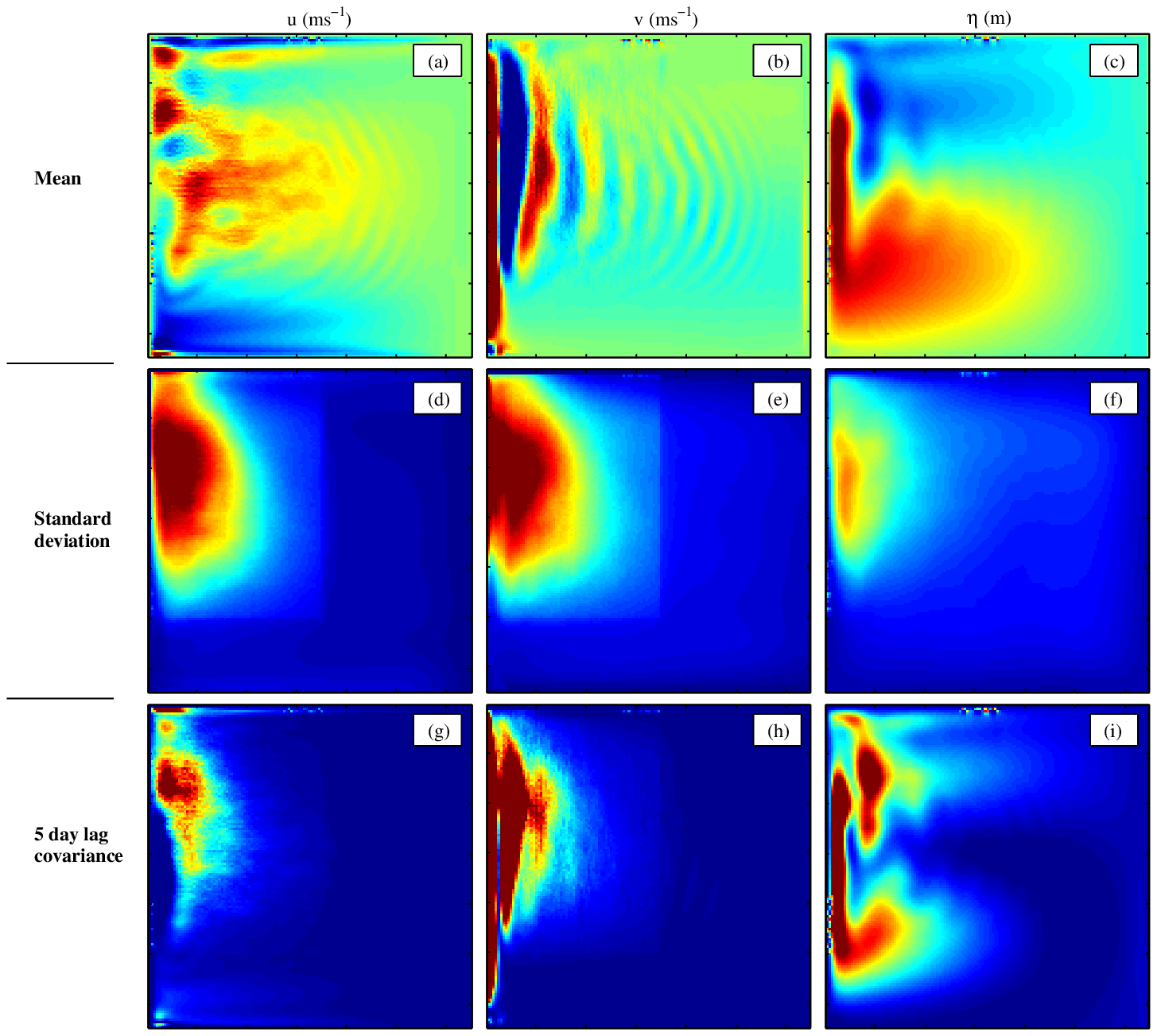}
\caption{Statistics of a low resolution integration with $F_u=\gamma_1 \nabla^2 \overline{u}_{\text{T}}$, $F_v=\gamma_1 \nabla^2 \overline{v}_{\text{T}}$, see the end of section \ref{constForce:sect}. In addition $q_u^{1/2}=q_v^{1/2}=0.004$ Pa s$^{-1/2}$ in the region of high variance and $b_u$ and $b_v$ are given by (\ref{guessStochForce:eqn}), see the end of section \ref{stochForce:sect}. Statistics of $u$ (left), $v$ (middle) and $\eta$ (right) including the mean (top), standard deviation (middle) and 5 day lag covariance (bottom) are plotted. To enable comparison, the colour axis for each plot is the same as that used in figures \ref{climMean:fig}, \ref{climVar:fig} and \ref{climLag5Var:fig}.}
\label{guessForcing:fig}
\end{figure*}

Comparing Figures \ref{paramVals:fig}b with \ref{paramVals:fig}c and \ref{paramVals:fig}e with \ref{paramVals:fig}f, indicates that the form of $\mathbf{q}_u$ and $\mathbf{q}_v$ is quite complex and is often relatively large in regions where the effective forcing ($\sqrt{-q_i/b_i}$) is small. We might therefore expect that the effective forcing is largely set by the time scale parameters $\mathbf{b}_u$ and $\mathbf{b}_v$. A simple approximate relation at each grid point is
\begin{equation}
  b_{ui}=\frac{\gamma_2}{\frac{1}{2} \rho_0 \left| \overline{u'^2_{i\text{T}}} - \overline{u'^2_{i\text{M}}}\right|}
  \qquad \text{and} \qquad
  b_{vi}=\frac{\gamma_2}{\frac{1}{2} \rho_0 \left| \overline{v'^2_{i\text{T}}} - \overline{v'^2_{i\text{M}}}\right|}
  \label{guessStochForce:eqn}
\end{equation}
where $\gamma_2$ is an undetermined constant, the subscripts M and T indicate data from the respective model and truth integrations, the over line indicates the time mean and the prime indicates the time varying part from the standard Reynolds decomposition, $u=\bar{u}+u'$ and $v=\bar{v}+v'$. As a test, we choose that $q_{ui}^{1/2}=q_{vi}^{1/2}=0.004$ Pa s$^{-1/2}$ in the region defined by the western and northern boundaries and 2100 km east and 2940 km south of these boundaries, $q_u=q_v=0$ elsewhere, $b_u$ and $b_v$ are given by (\ref{guessStochForce:eqn}) with $\gamma_2=2 \times 10^{-4}$ m$^2$ s$^{-3}$ and $F_u$ and $F_v$ are given by (\ref{guessConstForce:eqn}). The climatological statistics of the system integrated with these parameters are shown in Figure \ref{guessForcing:fig}. Comparing the plots in Figure \ref{climVar:fig} with \ref{guessForcing:fig} d, e and f, indicates that the variance of this test system has been improved relative to the low resolution model without parameterisation. It provides some evidence that the time scale of the sub grid forcing is important, rather than a particular spatial form of the amplitude. Unfortunately the 5 day lag covariance of the test system is poor, compare Figure \ref{climLag5Var:fig} with \ref{guessForcing:fig}, g, h and i. With long timescales in this test system, we would expect additional error in it's response to forcing.

\begin{table*}
\begin{tabular}{| p{2.4cm} | p{7cm} | p{7.2cm} |}
\hline
\textbf{Forcing terms used} & \textbf{Climatological mean} & \textbf{Climatological variance} \\ \hline
$F_u$
& The mean $u$ field converges relatively quickly, the $v$ and $\eta$ fields converge relatively slowly.
& No improvement in the amplitude or shape of the variance. \\ \hline
$F_v$
& The mean $v$ field converges relatively quickly, the $u$ and $\eta$ fields converge relatively quickly and then plateau at a relatively poor value of the mean squared error.
& No improvement in the amplitude or shape of the variance. \\ \hline
$F_u$, $F_v$
& The mean $u$, $v$ and $\eta$ fields converge relatively quickly.
& No improvement in the amplitude or shape of the variance. \\ \hline
$F_u$, $\xi_u$
& The mean $u$ field converges relatively quickly, the $v$ and $\eta$ fields converge relatively slowly.
& The variance in $u$ converges relatively quickly, and $v$ plateaus at a relatively poor value of the mean squared error. The 5 day lag covariance in $u$ and $v$ converges. No convergence in $\eta$. \\ \hline
$F_v$, $\xi_v$
& The mean $u$, $v$ and $\eta$ fields converge relatively quickly.
& The variance in $v$ converges relatively quickly, and $u$ plateaus at a relatively poor value of the mean squared error. The 5 day lag covariance in $u$ converges and $v$ converges relatively slowly. No convergence in $\eta$. \\ \hline
$\xi_u$, $\xi_v$
& No improvement in the amplitude or shape of the mean.
& The $u$ and $v$ variance fields converge relatively quickly. No convergence in $\eta$. \\ \hline
$d {\xi_u}_i=\sqrt{{q_u}_i} d{w_u}_i$ $d {\xi_v}_i=\sqrt{{q_v}_i} d{w_v}_i$ with various configurations of $F_u$ and $F_v$.
& The mean is optimised depending upon the inclusion of $F_u$ and $F_v$ as above.
& No improvement in the amplitude or shape of the variance. \\ \hline
$F_u$, $F_v$, $\xi_u$, $\xi_v$ (full parameter set)
& Fastest convergence and most accurate.
& Fastest convergence and most accurate. No convergence in $\eta$. \\
\hline
\end{tabular}
\caption{A summary of integrations performed using different
  combinations of parameterisation terms. The left column indicates
  the terms included and optimised for in equations (\ref{dudt:eqn}) and
  (\ref{dvdt:eqn}). Terms not included in the optimisation are equal to zero. The other two columns indicate the quality of the parameterised model climatological mean and variance as optimisation progresses.}
\label{optimiseWhat:table}
\end{table*}

\subsection{The response to a change in the wind}

Arguably, the most desirable property of a climate model, or a
seasonal forecast model, is an accurate response to changes in the
applied forcing. As mentioned in section \ref{ForcedRespIntro:sect}
and Figure \ref{changeAutoCorr:fig}, the fluctuation-dissipation
theorem suggests that a possible consequence of optimising the
timescales in the model is to improve the response to
forcing. Statistics related to the timescales, namely the variance and
5 day lag covariance, have been somewhat improved. In this section we
describe the impact upon the model's forced response.

\begin{figure*}
\includegraphics[width=\textwidth]{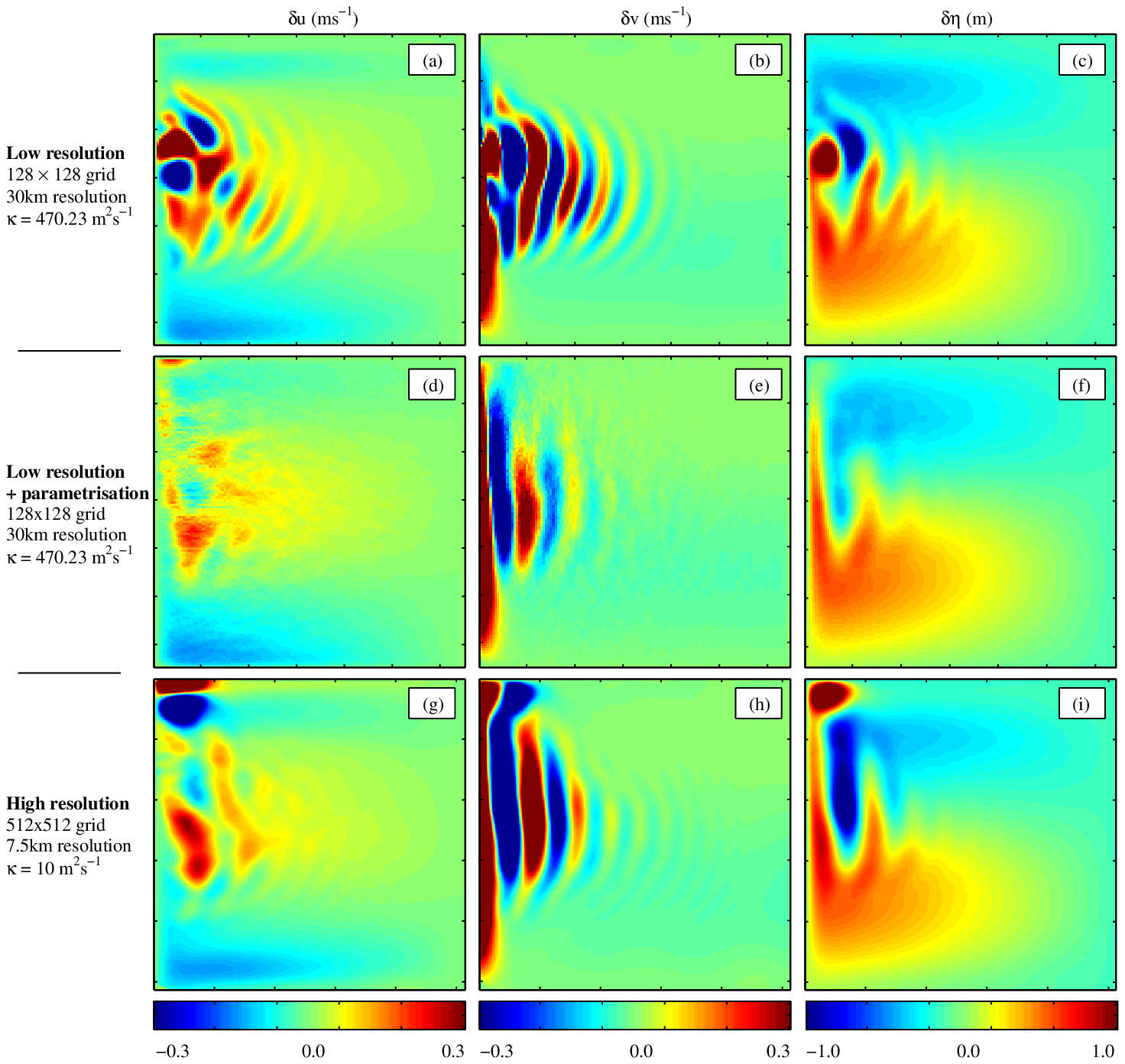}
\caption{The response to a small change to the wind forcing, in the
  zonal velocity $u$ (left), meridional velocity $v$ (middle) and sea
  surface height $\eta$ (right) for the low resolution (top), low
  resolution parameterised at iteration 150 (middle), and high
  resolution model (bottom). $\kappa$ is the viscosity parameter. The
  quantities shown are the ensemble mean climatological mean from the
  $\alpha_0=0.13$ Pa minus the ensemble mean climatological mean from
  the $\alpha_0=0.11$ Pa integrations.}
\label{forcedResponse:fig}
\end{figure*}

To test how the climatological mean of a particular configuration
responds to forcing, it is useful to apply several magnitudes of
forcing to check for linearity of the response, and to perform
multiple independent integrations with different initial conditions to
check for the uncertainty in the response. Therefore for five values
of the wind forcing amplitude $\alpha_0=0.10$, $0.11$, $0.12$, $0.13$
and $0.14$ Pa, an integration of ten ($N=10$) ensemble members with
random initial conditions was performed. Figure
\ref{forcedResponse:fig} shows the ensemble mean climatological mean
response of each system to a small change to the forcing. Uncertainty
in the response, quantified as the ensemble standard deviation of the
climatological mean multiplied by $2/\sqrt{N}$, peaked at around 10 to
15 percent in regions of large response. Given the variance in the
ensemble members, non-linearity of the response as a function of
forcing amplitude was undetectable.

Figure \ref{forcedResponse:fig} shows that the response to forcing of
the high resolution $u$ field (panel (g)) is dominated by the gyre in
the north-west corner of the domain (cf. Figure \ref{climMean:fig},
panel (g)). The low resolution $u$ field without parameterisation
(panel (a)) has a strong response close to the centre of the western
boundary. The parameterised system successfully reduces this response
to an amplitude more similar to the high resolution system but it
fails to introduce the response to the north-west corner. It is
plausible that the location of the circulation in the low resolution
unparameterised model is simply further north in the high resolution
model, and since our parameterisation is local it is not able to
reproduce such a non-local change. The response in the $v$ field is
located all along the western boundary in the high resolution system
(Figure \ref{forcedResponse:fig}, panel (h)), whereas the response in
the low resolution system without parameterisation (panel (b)) stops
half way and has a higher frequency pattern in the eastwards direction
close to the centre of the domain. The low resolution system with
parameterisation reproduces the pattern of the high resolution
response in $v$ quite well (compare panels (e) and (h)) although the
amplitude is lower. However in the north, once again the response is
not captured. A similar story applies to the $\eta$ field (panels (c),
(f) and (i)). The mean squared difference between the low resolution
parameterised and high resolution $u$, $v$ and $\eta$ fields is 0.606,
0.500 and 0.650 times the mean squared difference between the low
resolution unparameterised and high resolution fields
respectively. Values below one indicate that for each field, according
to this measure, the parameterisation has improved the response.

\section{Discussion}
\label{conclusions:sect}

We have successfully optimised a high dimensional spatially varying
linear stochastic parameterisation scheme for barotropic sub-grid eddy
turbulence. By finding the mean sub-grid eddy forcing term, our method
is successful at reproducing the climatological mean of a high
resolution idealised shallow water ocean gyre using low resolution
integrations. By finding the spatially varying amplitude and timescale
of a local stochastic sub-grid forcing term, our method is also
successful at reproducing the climatological variance and 5-day
lag-covariance of the velocity variables. The response to forcing of
the low resolution parameterised test system was significantly closer
to the true high resolution response (estimated using the mean squared
difference) than the default system without parameterisation. Only
data from one integration of the high resolution system was used for
optimisation, so by substituting sufficient measurements, or
reanalysis data, representing the real world, this method is
potentially useful in climate change experiments. The time evolution
of both the low and high resolution test systems is chaotic and we do
not foresee serious difficulties with application to more complex
systems.

The key to our method is optimisation. In a system with many
parameters it is impossible to explore the entire space defined by
them and a direction to optimise in must be assumed. This is normally
found using a tangent linear approach. We make the simpler assumption
that the direction to optimise in is given by the difference between
low and high resolution climatological states. When measurements or a
high resolution integration is available, optimisation allows us to
avoid complications such as having to fit the sub-grid model as in
\cite{Achatz99}, or spectral representations of the flow as in
\cite{frederiksenetal2006}. We also avoid making assumptions present
in less empirical theories that are either difficult to implement in
reality, (e.g. \citealp{kraichnan1959,Mana14}), or do not apply in
practical cases. The result is a simple and accurate method that can
be applied without modification to state of the art ocean models.

The measurements that we require are not too demanding. We require for
example the velocity measured at a single point in space at a
specified time. We do not require the tendencies that are used in
\cite{frederiksenetal2006}, \cite{Achatz99} and \cite{Achatz13}. We
require sufficient measurements to be able to estimate the
climatological mean, variance and lag-covariance in a region over some
period of time.

In pioneering work, using a tangent linear model and its adjoint,
\cite{ferreiraetal2005} optimise the eddy stresses of a $4^{\circ}
\times 4^{\circ}$ resolution global ocean model to obtain a model with
an accurate climatological mean temperature as defined by observations
\citep{Levitus-Boyer-1994:world}. As is the case with all sufficiently
coarse resolution ocean models, their global model was integrated with
a viscosity sufficient to damp away all of the chaotic eddies. In the
absence of a time varying forcing all time differentials are equal to
zero and their system is not chaotic. This enables the use of an
adjoint method to optimise the climatology. By contrast, the low
resolution model we use in this study is chaotic. Unfortunately, for
climate problems, adjoint methods like that applied by
\cite{ferreiraetal2005} cannot easily cope with chaotic systems
(see e.g. \citealp{Lea-Allen-Haine-2000:sensitivity} and
\citealp{Eyink04}). It is therefore difficult to extend them to higher
resolution models which are chaotic due to having lower viscosity.

Our simple approach requires many iterations (150 in our test case) of
a low resolution model integration. \cite{ferreiraetal2005} use 120
iterations of their forward and adjoint model. Given the different
model configurations, number of degrees of freedom and optimisation
tools, it remains difficult to assess which method is computationally
cheaper. We applied a basic Euler method at each iteration step so it
may be possible to reduce the number of iterations by using a higher
order approach. Additionally the direction that we push our
parameterisation vectors is not optimal, therefore it may also be
possible to reduce the number of iterations by more accurately
estimating this direction using a modified adjoint method
\citep{Wang14}, or by starting the optimisation with parameters
defined by a fit to the high resolution statistics as in
\cite{Achatz99} and \cite{frederiksenetal2006}, rather than the low
resolution climatology.

Since our shallow water ocean gyre test system exhibits a high degree
of geostrophic balance and is well approximated by the equivalent
single layer quasi-geostrophic system, one may expect that
optimisation of a single prognostic variable, $u$, $v$ or $\eta$,
would be sufficient. However this was found not to be the case. The
variability of the sea surface height was not forced, or well
reproduced in the low resolution model. Thus there is scope for
improvement by including a stochastic term in its governing
equation. The fluctuation-dissipation theorem guarantees that the
response to a forcing is related to the underlying variability, so
this may also improve the response of the parameterised model. When
estimating the forced response, we neglected any change in the
sub-grid parameters as a result of the forcing. So the low resolution
parameterised response estimate can potentially be improved by
including these changes as estimated using the fluctuation-dissipation
theorem \citep{Achatz13}. A complementary approach is to include
optimisation of the correlation between grid points of the stochastic
variables.


For objective measures of the quality of any model of a physical
system, measurements of the system to be represented are absolutely
necessary. In our case and in those of \cite{berloff2005},
\cite{Achatz99}, \cite{Achatz13}, \cite{frederiksenetal2006} and
\cite{Zidikheri09}, the measurements are represented by values taken
from a higher resolution integration. In this paper, in addition to
assessing the quality of our parameterisation, the high resolution
integration is used as an optimisation target. For a more realistic
ocean model, interpolated measurements of the real ocean, or
reanalysis data, would be used instead and the high resolution
integration is therefore not required. For example
\cite{ferreiraetal2005} use 1994 world ocean atlas data and more up to
date reanalysis is available for sectors of the ocean,
e.g. \cite{Mazloff10}. We have not developed a fully self-consistent
theory of ocean turbulence so it is impossible for us to derive the
values of our high dimensional parameterisation vector in advance. In
regions where we have no measurements (or theory), assumptions must be
made. For example we may assume that the parameters are all zero and
the default model is the best, or that parameters are the same as in a
similar region of ocean, or even that the parameters are given by
experiments with a high resolution regional model. Another approach is to assume some given form of the forcing. As an example we considered the hypothesis that the constant forcing is given by the viscous terms in the model, (multiplied by an unknown constant). In our model setup, a simple relation between the the eddy kinetic energy difference and the timescales of the stochastic forcing, was tested. Our results suggest that the impact of the stochastic forcing is mainly determined by the spatial pattern of these timescales rather than the spatial pattern of the amplitude of the variance. Using our method,
the numerical values of the parameters change if the model parameters
(density, resolution etc.) or boundary conditions (e.g. location of
continents) change. The solution to this problem, for those who wish
to make use of a particular model in a situation for which the
appropriate parameters have not been found, is to apply the
optimisation algorithm detailed here to find a new set of optimal
parameters.

Our scheme is very simple but makes use of three parameters and one
variable for each model grid cell. Although this large number of
parameters is not a practical disadvantage, it is not particularly
elegant. Through understanding of the ocean system
\citep{Gent-Mcwilliams-1990:isopycnal}, or turbulence in general
\citep{Holm99}, it may be possible to find a scheme of similar
quality using fewer parameters. Since we have been able to estimate
the mean sub-grid eddy forcing and have the amplitude and timescale of
a variable forcing term (in our case stochastic), our approach may
help with insights into a more developed theory. One advantage of our
method is that it does not conflict with other parameterisations. If
necessary it can be used in conjunction with the parameterisation
schemes already present in complex ocean models both at coarse or
eddy-permitting resolution.



\appendix

\section{Influence of a stochastic parameterisation upon a linear system}
\label{linearParam:sect}

Assume that a single element one dimensional model, $x$, is represented by a linear stochastic system of the form
\begin{equation}
dx=ax \, dt + \sqrt{b} \, dw
\label{linearNotParam:eqn}
\end{equation}
where $w$ represents Gaussian white noise with zero mean and unit variance and $a$ and $b$ are constants. We wish to add an additional term $r$ to the right hand side as a parameterisation giving
\begin{equation}
dx'=ax' \, dt + \sqrt{b} \, dw + r \, dt
\label{linearParam:eqn}
\end{equation}
where $r$ is given by
\[
dr=cr \, dt + \sqrt{g} \, dw'.
\]
$w'$ is also a white noise term with the same properties as $w$ and $c$ and $g$ are constants.

\begin{figure}
\begin{center}
\includegraphics[width=0.49\textwidth]{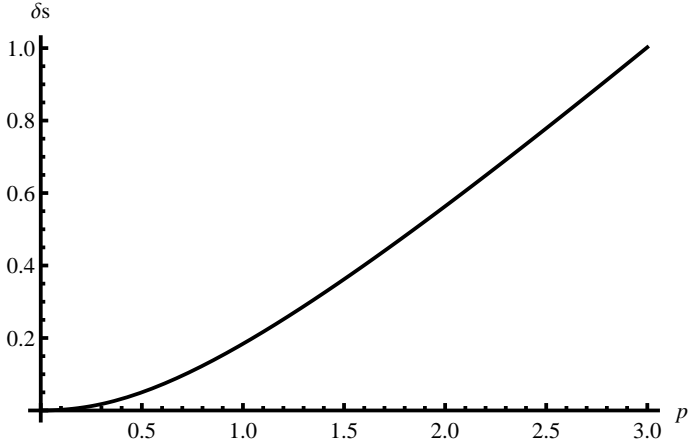}
\end{center}
\caption{The lag covariance $\delta s$ added to a linear model by a linear stochastic term with a timescale $p$ measured in units of time. Equation (\ref{lagVarDiffP:eqn}) with $a=-1$, $g=-1$ and $\tau =1$.}
\label{lagVarVsP:fig}
\end{figure}

Using the fact that for a $d$ dimensional linear system $\mathbf{C}(\tau)=\exp \left( \mathbf{B} \tau \right) \mathbf{C}(0)$ where $\mathbf{C}(\tau)$ is the systems lag $\tau$ covariance matrix, $\mathbf{B}$ is a constant matrix and we use the matrix exponential, it can be shown that the lag $\tau$ variance difference, $\delta s$, between (\ref{linearNotParam:eqn}) and (\ref{linearParam:eqn}) is given by
\begin{equation}
\delta s = \frac{cg\exp \left( a\tau \right) -ag\exp \left( c\tau \right)}{2a^3c+2ac^3}, \qquad c<0.
\label{lagVarDiff:eqn}
\end{equation}
Substituting the variable $p=-1/c$ into (\ref{lagVarDiff:eqn}) gives
\begin{equation}
\delta s = \frac{gp^2 \left( \exp \left( a \tau \right) + ap \exp \left( -\tau/p \right) \right)}{2a^3 p^2 - 2a}, \qquad p>0
\label{lagVarDiffP:eqn}
\end{equation}
Then over some region of $p+\delta p$, for $\tau >0$, $\delta s$ is approximately proportional to $p$, so changing $p$ in an iterative step (\ref{iterateB:eqn}) will lead to a reasonable change in the lag covariance, see Figure \ref{lagVarVsP:fig}.

\section*{Acknowledgements}
This work was funded by UK NERC grant NE/J00586X/1. We thank the three referees for their input which undoubtedly led to improvement of this paper. Thanks to David Munday and Miroslaw Andrejczuk for a huge amount of help with the MITgcm and thanks to James Maddison, Peter D\"{u}ben, PierGianLuca Porta Mana, Mark Forshaw, and Tim Palmer for the many discussions that contributed to the ideas in this paper.

\bibliographystyle{elsarticle-harv} 
\bibliography{all,portamanabib,references}





\end{document}